\documentclass[superscriptaddress,nofootinbib,preprintnumbers,amsmath,amssymb, preprint]{revtex4-1}
\usepackage[T1]{fontenc}

\usepackage{xpatch}
\makeatletter
\patchcmd{\@ssect@ltx}
    {\addcontentsline{toc}{#1}{\protect\numberline{}#8}}
    {}
    {}
    {}
\makeatother

\usepackage[titles]{tocloft}
\cftsetindents{section}{0em}{2.5em}
\cftsetindents{subsection}{2.5em}{2.5em}
\usepackage{multirow}
\usepackage{tabularx}
\usepackage{amsfonts}
\usepackage{mathrsfs}
\usepackage{leftidx}
\usepackage{amsmath}
\usepackage{amssymb}
\usepackage{placeins}
\usepackage{relsize}
\usepackage{slashed}
\usepackage{gensymb}
\usepackage{gensymb}
\usepackage{empheq}

\usepackage[dvipsnames]{xcolor}
\definecolor{red}{rgb}{0.9, 0,0}
\definecolor{cerulean}{rgb}{0., 0.42,0.9}
\definecolor{navy}{rgb}{0.05, 0.05,0.8}

\usepackage[colorlinks]{hyperref}
\hypersetup{
     colorlinks   = true,
     citecolor    = red,
	linkcolor = navy
}

\usepackage{soul}
\usepackage{epsfig}
\usepackage{graphicx}               
\usepackage{url}
\usepackage{float}
\usepackage{soul}

\newcommand{\be}{\begin{equation}}
\newcommand{\ee}{\end{equation}}
\newcommand{\bea}{\begin{eqnarray}}
\newcommand{\eea}{\end{eqnarray}}
\newcommand{\beq}{\begin{eqnarray}}
\newcommand{\eeq}{\end{eqnarray}}
\def\bit{\begin{itemize}}
\def\eit{\end{itemize}}
\def\ben{\begin{enumerate}}
\def\een{\end{enumerate}}
\newcommand{\Fig}[1]{Fig.~\ref{#1}}
\newcommand{\Eq}[1]{Eq.~(\ref{#1})}

\newcommand{\overbar}[1]{\mkern 2.5mu\overline{\mkern-2.5mu#1\mkern-2.5mu}\mkern 2.5mu}

\def\ln{\textrm{ln}}

\newcommand{\bfv}{{\bf v}}

\newcommand{\angstrom}{\mbox{\normalfont\AA}}

\newcommand{\mA}{m_{A'}}
\newcommand{\gD}{g_D}

\newcommand{\La}{\mathscr{L}}
\newcommand{\e}{\boldsymbol{\epsilon}}
\newcommand{\q}{\mathbf{q}}
\newcommand{\bk}{\mathbf{k}}
\newcommand{\K}{\mathcal{K}}
\newcommand{\Sm}{\mathcal{S}}

\newcommand\DN[1][\relax]{%
\ifx\relax#1\relax\else{}^{#1}\fi \!X}

\newcommand{\cm}{{\,\rm cm}}
\newcommand{\g}{{\,\rm g}}
\newcommand{\Kg}{{\,\rm kg}}
\newcommand{\Km}{{\,\rm Km}}
\newcommand{\s}{{\,\rm s}}

\newcommand{\keV}{{\,\rm keV}}
\newcommand{\eV}{{\,\rm eV}}
\newcommand{\meV}{{\,\rm meV}}

\DeclareMathAlphabet\mathbfcal{OMS}{cmsy}{b}{n} 

\def\eq#1{Eq.~(\ref{#1})}

\graphicspath{{./figures/}}

\definecolor{cerulean}{rgb}{0., 0.52,0.65}

\begin{document}


\title{Directional Dark Matter Detection in Anisotropic Dirac Materials}

\author{Ahmet Coskuner}
\affiliation{Berkeley Center for Theoretical Physics, University of California, Berkeley, CA 94720, USA}
\affiliation{Department of Physics, University of California, Berkeley, CA 94720, USA}
\author{Andrea Mitridate}
\affiliation{Scuola Normale Superiore, Piazza dei Cavalieri 7, 56126, Pisa, Italy}
\affiliation{INFN sezione di Pisa, Italy}
\affiliation{Walter Burke Institute for Theoretical Physics,
California Institute of Technology, Pasadena, CA 91125}
\author{Andres Olivares} 
\affiliation{Institute for Particle Physics Phenomenology, Department of Physics, Durham University,
South Road, Durham DH1 3LE, United Kingdom}
\author{Kathryn M. Zurek}
\affiliation{Walter Burke Institute for Theoretical Physics,
California Institute of Technology, Pasadena, CA 91125}

\addvspace{4ex}
\begin{abstract}
Dirac materials, because of their small ${\cal O}(\mbox{meV})$ band gap, are a promising target for dark photon-mediated scattering and absorption of light dark matter.  In this paper, we characterize the daily modulation rate of dark matter interacting with a Dirac material due to anisotropies in their crystal structure. We show that daily modulation is an ${\cal O}(1)$ fraction of the total rate for dark matter scattering in the Dirac material ZrTe$_5$. When present, the modulation is dominated by the orientation of the material's dielectric tensor with respect to the dark matter wind, and is maximized when the crystal is oriented such that the dark matter wind is completely aligned with the largest and smallest components of the dielectric tensor at two different times of the day.  Because of the large modulation, any putative dark matter scattering signal could be rapidly verified or ruled out by changing the orientation of the crystal with respect to the wind and observing how the daily modulation pattern changes.
\end{abstract}

\maketitle

\tableofcontents


\section{Introduction}

The direct detection of dark matter has seen a dramatic broadening of scope beyond the traditional domains of the search for the Weakly Interacting Massive Particle (WIMP) and the axion.  Both are highly motivated by solving other problems in the SM sector, such as the hierarchy problem (WIMP) and the strong CP problem (axion).  However, the WIMP has been increasingly constrained by highly sensitive experiments that have searched for new physics at the weak scale for decades. 

At the same time, visible and dark matter may originate from two separate sectors, independent of any problems in the SM \cite{Strassler:2006im}.  The development of the hidden sector paradigm has opened a fruitful direction in building models of dark matter.  Similar to the SM, the lightest particle of the dark sector may be stable thanks to a gauge or  global symmetry. The interactions with the SM may be only gravitational, or mediated by new forces that interact only very feebly with the dark and/or visible sectors.  However, these feeble interactions with the SM may be crucial in setting the dark matter abundance, as is typical, for example, in models of MeV dark matter \cite{Boehm:2003hm,Pospelov:2007mp,Hooper:2008im}, WIMPless miracle DM \cite{Kumar:2009bw}, Asymmetric DM \cite{Kaplan:2009ag,Cohen:2010kn}, GeV hidden sector dark matter \cite{ArkaniHamed:2008qp,Cheung:2009qd,Morrissey:2009ur}, freeze-in DM \cite{Hall:2009bx}, and Strongly Interacting Massive Particles \cite{Hochberg:2014dra}.  In this case, well-motivated and clear benchmarks exist to search for dark matter in hidden sectors.  Two classic examples are Asymmetric Dark Matter annihilating its symmetric abundance through the dark photon \cite{Lin:2011gj}, as well as dark-photon mediated dark matter produced through a freeze-in process \cite{Chu:2011be,Dvorkin:2019zdi}. 

While there are many well-motivated models, if the mass of the dark matter in the hidden sector is below approximately a GeV, existing technologies to search for the WIMP through nuclear recoils are not sensitive to it.  However, recent years have seen the development of many new ideas to detect sub-GeV dark matter (see Ref. \cite{Battaglieri:2017aum} for an extensive review). When the dark matter carries more energy than electronic excitation energies--typically in the 1-10 eV range, corresponding to the kinetic energy of 1-10 MeV dark matter--semiconductor \cite{Essig:2011nj,Graham:2012su,Essig:2015cda} or noble liquid \cite{Essig:2012yx} targets developed to search for the WIMP through nuclear recoils with keV energy deposition can be extended to search instead for valence-to-conduction-band transitions or ionization.  The challenge is to reduce dark counts and increase sensitivity to energy depositions three orders of magnitude below that needed for WIMPs.  This program is well under way in noble liquid \cite{Essig:2017kqs} and silicon targets, and is actively being pursued in collaborations such as SENSEI \cite{Crisler:2018gci,Tiffenberg:2017aac,Abramoff:2019dfb}, DAMIC \cite{Aguilar-Arevalo:2016ndq,Aguilar-Arevalo:2019wdi} and SuperCDMS \cite{Agnese:2014aze,Agnese:2015nto,Romani:2017iwi,Agnese:2018col,Agnese:2018gze}.

When the DM mass drops below an MeV, new ideas and targets must be found.  This has been the focus of intense efforts on small gap materials, such as superconductors \cite{Hochberg:2015pha,Hochberg:2015fth}, graphene \cite{Hochberg:2016ntt}, and Dirac Materials \cite{Hochberg:2017wce} in the case of electronic excitations, and superfluid helium \cite{Knapen:2016cue,Schutz:2016tid} and polar materials \cite{Knapen:2017ekk} when the DM couples to nuclei or ions.  

In this paper, we will focus on electronic excitations, and in particular dark photon mediated couplings to electrons in Dirac semimetals. These materials have a small band gap of order $\mathcal{O}(1-10~\meV)$, allowing to probe keV-$\rm MeV$ mass DM through scattering. 
Superconductors are another small gap material that can probe sub-MeV dark matter interacting with the electromagnetic current via a kinetically mixed dark photon, but in-medium effects have been shown to have a dramatic effect on the dark matter interaction rate \cite{Hochberg:2015fth}.  The large in-medium effects (or, equivalently, optical response) are a result of the large electron density of states near the Fermi surface of a metal.  
Dirac materials were proposed in Ref.~\cite{Hochberg:2017wce} as an antidote to this problem: the optical response of Dirac materials is much smaller than superconductors and more similar to a semiconductor.  At the same time, the presence of states within  $\mathcal{O}(1-10~\meV)$ of the Fermi surface gives sensitivity to similarly light DM as for superconductors.  An additional desirable feature is that Dirac materials feature anisotropic Fermi velocities and permittivity tensor, suggesting that dark matter interaction rates should be dependent on the orientation of the material with respect to the dark matter wind, inducing a daily modulation.  Such a modulation is a smoking gun signature for dark matter interactions, distinguishing signal from background. 

The goal of this paper is to explore in detail how direct detection signals are affected by Dirac materials anisotropies: both the electron Fermi velocities, which dominate the kinematics of the DM-electron scattering, and the permittivity tensor, which characterizes the strength of the electron interactions with the dark photon.  We find that daily modulation is ${\cal O}(1)$ for DM scatterings, while no modulation is present for DM absorption. Since the scattering rate modulation turns out to be dominated by anisotropies in the crystal dielectric, it can be maximized orienting the crystal in such a way that the DM wind is completely aligned with the smallest and largest component of the dielectric tensor at two different times of the day.

Strong anisotropy and daily modulation in sub-MeV dark matter interaction rates are also a feature of polar materials, as proposed in Ref.~\cite{Griffin:2018bjn}, though in Dirac Materials the interaction is with electrons rather than ions.  Other proposals for directional detection of DM candidates include using two-dimensional materials like graphene \cite{Hochberg:2016ntt} in the case of DM coupling to electrons; defect production in crystals \cite{Budnik:2017sbu,Rajendran:2017ynw}, and nuclear recoils in semiconductors \cite{PhysRevLett.120.111301}.

The outline of this paper is as follows.  In the next Section we lay out the formalism for calculating dark matter interactions with in-medium effects.  In Sec.~\ref{sec:DMwind}, we define our conventions for the orientation of the crystal with respect to the DM wind.  Then in Secs.~\ref{sec:abs} and \ref{sec:scatt}, we compute the anisotropic absorption and scattering rates.  Finally we conclude.

 
 \section{Dark Matter Interactions with In-Medium Effects} \label{sec:inmediumformalism}
  
As a benchmark model we consider a dark sector coupled to the Standard Model through a kinetically mixed dark photon,
 \be\label{vacuumLa}
 \La \supset -\frac{1}{4}F_{\mu\nu}F^{\mu\nu}-\frac{1}{4}F_{\mu\nu}'F'^{\mu\nu}+\frac{\varepsilon}{2}F_{\mu\nu}F'^{\mu\nu}+eJ^\mu_{\rm EM}A_\mu+g_{\rm D} J^\mu_{\rm DM}A'_\mu+\frac{\mA^2}{2}A'^\mu A'_\mu\,,
 \ee
 where $F_{\mu\nu}$ ($F_{\mu\nu}'$) is the electromagnetic (dark) field strength, $\varepsilon$ the kinetic mixing parameter and $J^\mu_{\rm EM\;(DM)}$ the electromagnetic (dark) current that couples to (dark) photons with a coupling $e$ ($\gD$). The dark photon has a mass $\mA$ which can be generated either by a dark Higgs or through the Stueckelberg mechanism (though astrophysical constraints are weakest for the case of a Stueckelberg dark photon). In the vacuum, the propagating photon (found by diagonalizing the kinetic term in \eq{vacuumLa}) is $\tilde A_\mu=A_\mu-\varepsilon A_\mu'$. In this basis the dark photon mass eigenstate $A'$ couples to the electromagnetic current with a strength $e\varepsilon$. 
 
 Because of this coupling, the propagation of a dark photon in optically responsive media is modified. Specifically, including in-medium effects, we can write down the effective Lagrangian (in the $\tilde A$, $A'$ basis) 
 \begin{align}\label{inmediumLa1}
 \La\supset& -\frac{1}{4}\tilde F_{\mu\nu}\tilde F^{\mu\nu}-\frac{1}{4}F_{\mu\nu}'F'^{\mu\nu} +eJ^\mu_{\rm EM}\left(\tilde A_\mu+\varepsilon A'_\mu\right)+g_{\rm D} J^\mu_{\rm DM}A'_\mu+\frac{\mA^2}{2}A'^\mu A'_\mu\\
 &+\frac{1}{2}\tilde A^\mu\Pi_{\mu\nu}\tilde A^\nu+\varepsilon \tilde A^\mu\Pi_{\mu\nu}A'^\nu
 \end{align}
 where the in-medium polarization tensor is defined as $\Pi_{\mu\nu}\equiv i e^2 \langle J^\mu_{\rm EM}J^\nu_{\rm EM}\rangle$, and the equation of motion has been used to derive the last line. In the familiar case of isotropic materials, the polarization tensor can be written as $\Pi^{\mu\nu}=\Pi_T (\epsilon_+^{\mu}\epsilon_+^{\nu}+\epsilon_-^{\mu}\epsilon_-^{\nu})+\Pi_L\epsilon^{\mu}_L\epsilon^{\nu}_L$, with $\epsilon_L$ and $\epsilon_{+,-}$ the longitudinal and transverse polarization vectors. For $q\parallel\hat z$, they can be written as:
\be
 \epsilon_L = \frac{1}{\sqrt{q^2}}\Big(|\q|,\omega\,\hat q\Big)\qquad\epsilon_\pm=\frac{1}{\sqrt{2}}\Big(0,1,\pm i,0\Big)\,.
\ee 
One then finds that transverse and longitudinal dark photons remain decoupled during their propagation in the medium and interact with the electromagnetic current with reduced couplings \cite{Hochberg:2015fth}:
 \be
 \La\supset \varepsilon e\frac{q^2}{q^2-\Pi_{L,T}}A^{'T,L}_\mu J^\mu_{\rm EM}\,.
 \ee
 
 In the case of anisotropic materials, the polarization tensor cannot be decomposed into a longitudinal and a transverse component. This induces a mixing between longitudinal and transverse polarizations that can be parametrized in terms of a symmetric $3\times3$ mixing matrix, $\K$, defined as
 \be\label{eq:mixing}
 \K_{AB}\equiv(\epsilon_A^{\mu})^\star\Pi_{\mu\nu}\epsilon_B^\nu,
 \ee
 with $A$ and $B$ running over longitudinal and transverse polarizations. It is therefore useful to choose a basis, $\epsilon_{i=1,2,3}^\mu$, for the physical polarizations that is not mixed by in-medium propagation. This basis is found using the $3\times3$ unitary matrix, $\Sm$, that diagonalizes the mixing matrix:
 \be\label{eq:propbasis}
 (\epsilon_1^{\,\mu},\epsilon_2^{\,\mu},\epsilon_3^{\,\mu})=\Sm\left(\begin{array}{c} \epsilon_L^{\,\mu}\\\epsilon_+^{\,\mu}\\\epsilon_-^{\,\mu}\end{array}\right)\quad\quad \textrm{with}\quad\quad \Sm^{-1}\K\Sm={\rm diag}\Big(\pi_1(q),\pi_2(q),\pi_3(q)\Big)\,.
 \ee
In this new basis, the Lagrangian of \eq{inmediumLa1} takes the form 
 \begin{align}\label{inmediumLa2}
 \La\supset & -\frac{1}{4}\tilde F_{\mu\nu}^i\tilde F^{\mu\nu}_i-\frac{1}{4}F_{\mu\nu}^{'i}F'^{\mu\nu}_i+eJ^\mu_{\rm EM}\left(\tilde A_\mu^i+\varepsilon A^{'i}_\mu\right)+g_{\rm D} J^\mu_{\rm DM}A^{'i}_\mu-\frac{\mA^2}{2}A^{'\mu}_i A^{'i}_\mu\\
 &-\frac{\pi_i}{2}\tilde A^{i}_\mu \tilde A^{\mu}_i-\varepsilon\pi_i \tilde A^{i}_\mu A^{'\mu}_i
 \end{align}
 and can be diagonalized making the following field redefinition:
  \be
 \tilde A_\mu^i= \bar A_\mu^i+\frac{\varepsilon \pi_i(q)}{\mA^2-\pi_i(q)} \bar A^{'i}_\mu \quad\quad  A_\mu^{'i}=  \bar A_\mu^{'i}-\frac{\varepsilon \pi_i(q)}{\mA^2-\pi_i(q)}\bar A^{i}_\mu\,.
 \ee
 One then finds that the propagating dark photons $\bar A'_i$ couple to the electromagnetic current as:
 \be\label{mixing}
 \La\supset \varepsilon e\frac{q^2}{q^2-\pi_i}\bar A^{'i}_\mu J^\mu_{\rm EM}\,.
 \ee 
Thus we see that in the anisotropic case the role of $\Pi_{L,T}$ is played by the mixing matrix eigenvalues $\pi_i$. 

 Working in Lorentz gauge, the in-medium photon propagator takes the form 
 \be\label{eq:prop}
 G^{\mu\nu}_{\rm med}(q)=\sum_{i}\frac{P_i^{\mu\nu}}{\pi_i-q^2}\quad 
 \ee
 where $q=(\omega,\q)$ is the four-momentum transfer and $P_i$ is the projector operator on the direction of the $i$-th polarization. In order to compute the eigenvalues of the mixing matrix, we relate its components to the optical properties of the medium (see appendix \ref{app:poltens} for details): 
 \be\label{eq:Kcomp}
 \K_{LL}=q^2\left(1-\hat\q\cdot\e\cdot\hat\q\right)\quad\quad \K_{L\pm}=-\omega q\,\hat q\cdot\e\cdot\hat \epsilon_\pm\quad\quad \K_{\pm\pm}=\omega^2\left(1-\hat \epsilon_\mp\cdot\e\cdot\hat \epsilon_\pm\right)\,
 \ee
 where $\e$ is the dielectric tensor.\footnote{We will neglect the ion contribution to the dielectric tensor since it is expected to be sub-leading compared to the electronic one.}
 
From \eq{eq:Kcomp}, it is clear that in the scattering limit ($|q^2|\sim\q^2\gg\omega^2$) the mixing matrix is dominated by the $\K_{LL}$ component. Therefore, for DM scattering there is a negligible mixing between longitudinal and transverse polarizations, and the rate is dominated by the longitudinal degrees of freedom whose in-medium propagator is given by
\be
G^{\mu\nu}=\frac{P_L^{\mu\nu}}{q^2(\hat\q\cdot\e\cdot\hat\q)}.
\label{eq:propagator}
\ee

Conversely, in the absorption limit, all components of the mixing matrix are of the same order. Hence in-medium propagation gives rise to a sizable mixing. Therefore, to study DM absorption we need to work in the basis defined in \eq{eq:propbasis} and use the general in-medium propagator given in \eq{eq:prop}. 
 
 \section{Preliminaries: Dark Matter Wind}\label{sec:DMwind}
 
\begin{figure}[t]
\centering
\includegraphics[width=0.65\textwidth]{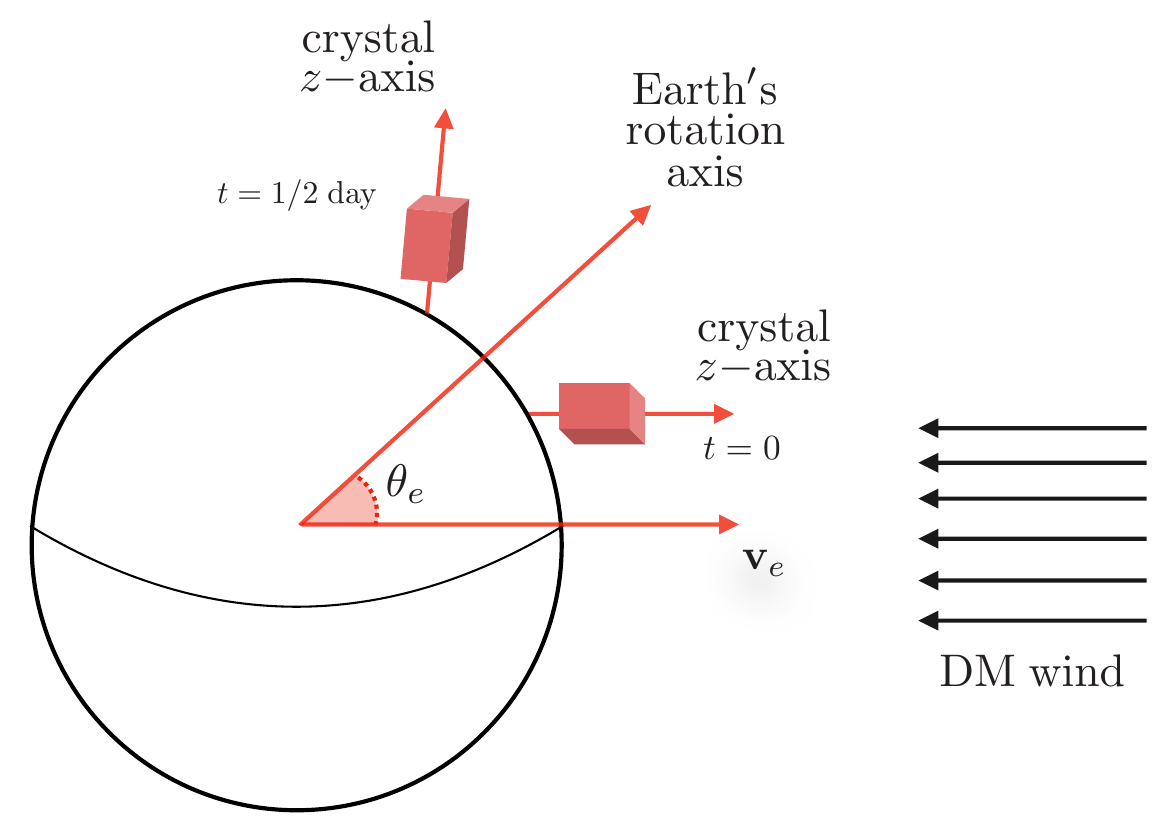}
\caption{\emph{Orientation setup of the experimental apparatus. At $t=0$ the $z$-axis of the crystal is aligned with the Earth's velocity (which is approximately in the direction of Cygnus). With this choice for the crystal orientation, the modulation of the signal is independent of the position of the laboratory.}}
\label{fig:orientation}
\end{figure}
 
The effects we are considering arise because the DM velocity with respect to the detector changes as the Earth rotates around its axis, see Fig.~\ref{fig:orientation}.  There are two kinds of effects.  The first is kinematic, where the flux of DM particles in the kinematic configuration that can excite a response in the target changes on a daily basis.  This effect is dominated by the anisotropy of the Fermi velocity in the material.  The second is due to the size of the matrix element, as shown for example in \eq{eq:propagator}, where the direction of the momentum transfer (typically oriented along the DM wind) changes with respect to the anisotropic dielectric tensor $\e$.  We will detail these effects separately for both absorption and scattering below.  

But before moving to the results we summarize here our conventions for the DM velocity distribution and the orientation of the DM wind in the crystal rest frame, using a set-up similar to Ref.~\cite{Griffin:2018bjn}.  For the velocity distribution in the galactic rest frame, $f_{\rm gal}(\bfv)$, we assume a Maxwellian form, with velocity dispersion $v_0=220\Km/\s$, truncated at $v_{esc}=500\Km/\s$. The velocity distribution in the laboratory frame is related to the one in the galactic frame by $f_{\rm lab}(\bfv,t)=f_{\rm gal}(\bfv+\bfv_e(t))$:
\be
f_{\rm lab}(\bfv,t)= \frac{1}{N_0}  \exp \left[ - \frac{(\bfv + \bfv_e)^2}{v_0^2} \right]   \ \ \Theta( v_{\text{esc}} - |\bfv + \bfv_e|)\,, \label{eq:velodist} 
\ee
where ${\bf v}_e(t)$ is the Earth's velocity with respect to the DM rest frame due to its revolution around the Sun and $N_0$ is a normalization constant given by
\be\label{eq:normcost}
N_0 = \pi^{3/2} v_0^3 \left[ {\rm erf} ( \tfrac{v_{\text{esc}}}{v_0} ) - \frac{2}{\sqrt{\pi}}\frac{v_{\text{esc}}}{v_0} \exp\left( -( \frac{v_{\text{esc}}}{v_0} )^2 \right) \right]\,.
\ee
The net effect of the Earth's revolution is to induce a DM wind oriented in the opposite direction of $\bfv_e$ in the laboratory reference frame.  The orientation of ${\bf v}_e$, and therefore of the DM wind, relative to the crystal changes due to the rotation of the Earth. Choosing the crystal orientation such that at $t=0$ the z-axis in the crystal frame is aligned with the Earth's velocity the explicit form of ${\bf v}_e$ is:
\be
{\bf v}_e=|{\bf v}_e|
\left(
\begin{array}{c}
\sin\theta_e \sin \phi\\
\sin\theta_e\cos\theta_e\left(\cos\phi-1\right)\\
\cos^2\theta_e+\sin^2\theta_e\cos\phi
\end{array}
\right)
\ee
where $|{\bf v}_e|\approx 240\Km/\s$, $\phi=2\pi\times t/24{\,\rm h}$ is the angle parametrizing the rotation of the Earth around its axis and $\theta_e\approx 42^\circ$ is the angle between the Earth's rotation axis and the direction of its velocity (see Fig. \ref{fig:orientation} for an illustration of the orientation setup).  
 
 
 \begin{table}[t]
\begin{tabular}{ccc}
\hline\hline
Parameter                                   & \quad Value for ZrTe$_5$ (th.)    \quad   & \quad   Value for ZrTe$_5$ (exp.) \quad             \\ \hline \hline
$\textrm{Re}(\e_{xx},\,\e_{yy},\,\e_{zz})$ & $(187.5,\,9.8,\,90.9)$   \\
$(v_{F,x},\,v_{F,y},\,v_{F,z})$             & $(2.9,\,0.5,\,2.1)\cdot10^{-3}$ & $(1.3,\,0.65,\,1.6)\cdot10^{-3}$  \\ 
$\Delta$                                    & $2.5\meV$             & $11.75\meV$              \\ 
$\Lambda$                                   & $0.14\keV$                         \\
$g$                                         & $4$                                \\
$V_{\rm uc}$                                & $795 \angstrom^3$                  \\
$\rho_T$                                    & $6.1\g/\cm^3$                      \\
$n_e$                                       & $8.3 \times 10^{23} \ e^-/\rm{kg}$ \\\hline\hline
\end{tabular}
\caption{\emph{Theoretical and experimental values of the material parameters for $\textrm{ZrTe}_5$ \cite{Hochberg:2017wce}. Theoretical (experimental) values are used for the curves labeled ``th'' (``exp'') in plots throughout this paper. For the parameters that were not obtained experimentally, theoretical values are used for both cases. }}\label{tab:zrte5values}
\end{table}
 
 \section{Absorption in Anisotropic Dirac Materials}\label{sec:abs}
  
 The DM absorption rate per unit time and detector mass is given by:
 \be 
 R_{abs}=\int d^3 {\bf v}\, f_{\rm lab}({\bf v},t)R_{abs}(\omega,\q)
 \ee
 where $f_{\rm lab}({\bf v},t)$ is the DM velocity distribution in the laboratory frame, as defined in the previous section, and $R_{abs}(\omega,\q)$ is the DM absorption rate at fixed energy and momentum transfer:
 \be
 R_{abs}(\omega,\q)=\frac{1}{\rho_T}\frac{\rho_\chi}{m_\chi}\langle n_T\sigma_{\rm abs}v_{\rm rel}\rangle_{\rm DM}\,,
 \ee
where $\rho_T$ is the mass density of the target, $n_T$ the number of target particles, $\sigma_{\rm abs}(\omega,\q)$ the cross section for DM absorption by the target and $v_{\rm rel}$ the relative velocity between the target and the DM. For DM absorption, the energy transfer is uniquely fixed by the DM mass ($\omega=m_\chi$) and the momentum transfer is equal to the DM momentum ($\q=m_\chi{\bf v}$).
 
 To derive the absorption rate of dark photons we start by considering the one of ordinary photons. This is related to the optical properties of the medium through the optical theorem:
 \be 
 \langle n_T\sigma_{\rm abs}v_{\rm rel}\rangle_{\gamma_i}=\frac{1}{\omega}{\rm Im}\Big[(\epsilon_i^{\,\mu})^*\,\Pi_{\mu\nu}\,\epsilon_i^\nu\Big]=-\frac{1}{\omega}{\rm Im}\Big[\pi_i(q)\Big]
 \ee
 where $\epsilon_i^\mu$ are the polarization vectors that diagonalize the polarization tensor and $\pi_i(q)$ the corresponding eigenvalues (as discussed in Sec.~\ref{sec:inmediumformalism}). From \eq{mixing}, we see that the effective mixing between ordinary and dark photons is given by
 \be\label{eq:eeff}
 \varepsilon_{{\rm eff},i}^2=\frac{\varepsilon^2\mA^4}{\left[\mA^2-{\rm Re}\,\pi_i(q)\right]^2+\left[{\rm Im}\,\pi_i(q)\right]^2}
 \ee
 and so the dark photon absorption rate at fixed energy and momentum transfer reads
 \be\label{eq:rateq}
 R_{\rm abs}^{A'}(\omega,\q)=-\frac{1}{3}\frac{\rho_\chi}{\rho_T}\sum_{i=1}^3\,\varepsilon^2_{{\rm eff},i}\frac{{\rm Im}\pi_i(q)}{\omega^2}\,,
 \ee
 where the $1/3$ comes from the average over dark photon polarizations. The functional form of $\pi_i$, for the absorption limit ($\omega\gg|\q|$), can be computed for a specific orientation of $\q$. This is possible because the mixing matrices, $\K$ and $\K'$, for two different orientations of the momentum transfer, $\q$ and $\q'$, are related by a similarity transformation (\emph{i.e.} $\K'=R^T\K R$ where $R$ is a 3-by-3 matrix such that $\q'=R\q$) and therefore have the same eigenvalues. Hence, choosing for example $\q\parallel \hat z$, we find that the functional form of the three $\pi_i$ for the absorption case is given by:
\be
\pi_i = \omega^2(1-\e_{ii})\,,
\ee
which allows us to write the absorption rate as 
\be
  R_{\rm abs}^{A'}(\omega,\q)=\frac{\varepsilon^2}{3}\frac{\rho_\chi}{\rho_T}\sum_{i=1}^3\frac{{\rm Im}[\e_{ii}]}{\textrm{Re}[\e_{ii}]^2+\textrm{Im}[\e_{ii}]^2}\,.
\ee

\begin{figure}[t]
\begin{center}
\includegraphics[width=0.73\textwidth]{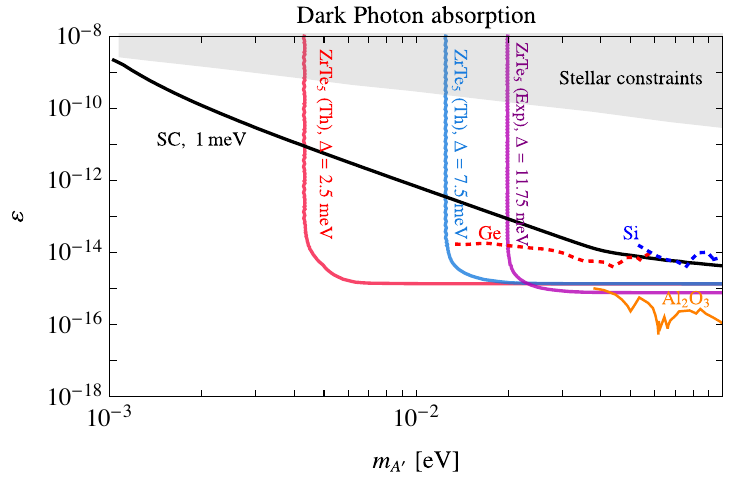}
\caption{\emph{Projected reach for the absorption of a kinetically mixed dark photon with a kinetic mixing parameter $\varepsilon$ and a mass $m_{A'}$. We show the $95\%$ C.L. sensitivity ($3$ events) that can be obtained with $1 \Kg$-year exposure of a Dirac material with the (where available, experimental) properties of ${\rm ZrTe}_5$ shown in Table~\ref{tab:zrte5values} but with a band gap of $2.5\meV$ (solid red), $7.5\meV$ (solid blue) and $11.75\meV$ (solid purple). For comparison we also show the sensitivity of a superconductor with a $1 \meV$ threshold (black line) \cite{Hochberg:2015fth}, sapphire ($Al_2O_3$) with a $1\meV$ threshold (orange) taken as reference for polar materials \cite{Griffin:2018bjn}, and two-phonon excitations in germanium (dotted red) and silicon (dotted blue) semiconductors \cite{Hochberg:2016sqx}. The gray shaded region is excluded by stellar emission constraints \cite{An:2014twa,An:2013yua}. The plot is cut at $m_{A'}=160\meV$, which is the largest energy deposit consistent with the linear energy dispersion relation.}}
\label{fig:absreachplot}
\end{center}
\end{figure}
As discussed in Appendix \ref{app:diel}, since for $\omega\lesssim\Lambda v_{F}$ the imaginary part of the dielectric receives contributions only from excitations close to the Dirac point, $\textrm{Im}[\e]$ can be computed using the analytic electron wave functions provided in \cite{Hochberg:2017wce}. For what concerns the real part, we need to resort to a DFT calculation whose results are shown in Table \ref{tab:zrte5values}.  Note this is in distinction to Refs.~\cite{Hochberg:2017wce} and \cite{Geilhufe:2019ndy} which used an analytic calculation of the real part of $\e$ to determine the rate.  This underestimates $\e$ because the analytic calculation, including only states near the Dirac point, does not give the contribution of the full electronic response.  In \Fig{fig:absreachplot} we show the projected reach for a ${\rm ZrTe}_5$ target with a gap of $2.5\meV$ and $7.5\meV$, and we compare it with the reach of polar materials \cite{Griffin:2018bjn}, superconducting \cite{Hochberg:2015fth} and semiconducting targets \cite{Hochberg:2016sqx}. The Dirac material reach shown here is almost one order of magnitude smaller than previously claimed in references \cite{Hochberg:2017wce} and \cite{Geilhufe:2019ndy} because of their underestimate of $\e$.  We summarize in more detail in Appendix~C the differences between the present calculation and those that appeared previously.


 \section{Scattering in Anisotropic Dirac Materials}\label{sec:scatt}

We will characterize the dark matter interaction rate in terms of the target response (the so-called dynamic structure factor $S(q,\omega)$), the dark matter velocity phase space characterized by the DM velocity distribution $ f_{\textrm{\rm gal}} (\vec{v})$, and the nature of the mediator (massless or massive).  Then we have the rate to scatter from the valence band of the crystal ($-$, momentum $\bk$) to the conduction band ($+$, momentum $\bk'$) given by \cite{Essig:2015cda}
\be
R_{\bf{k},\bf{k'}} = \frac{1}{\rho_T}\frac{\rho_\chi}{m_\chi} \frac{\pi\overbar{\sigma}_e}{ \mu_{\chi e}^2} \int \frac{d^3 \textbf{q}}{(2\pi)^3} \int d^3 {\bf v} f_{\textrm{\rm lab}} ({\bf v})   \  F_{\textrm{med}}(q)^2 S({\bf q},\omega), \\
\ee
where $\mu_{\chi e}$ is the DM-electron reduced mass, 
\beq
\omega = \frac{1}{2} m_\chi v^2 - \frac{(m_\chi {\bf v} - {\bf q})^2}{2 m_\chi} = {\bf q } \cdot {\bf v} - \frac{\q^2}{2 m_\chi}
\eeq
is the energy deposition, and the dynamic structure factor  $S({\bf q},\omega)$  is 
\be
S(\textbf{q},\omega) =2\pi  \left(\frac{{\bf q}^2}{|\textbf{q} \cdot \boldsymbol{\epsilon} \cdot \textbf{q}|}\right)^2 \frac{1}{2}\frac{(2 \pi)^3}{V}  \left(1 -  \frac{\tilde{\bk} \cdot \tilde{\bf{k'}}}{|\tilde{\bf{k}}| |\tilde{\bf{k}'}|} \right) \delta^3(\textbf{q} - (\bk' - \bk)) \delta(\Delta E_{\bk ,\bk'} - \omega )\,,
\ee
where $V$ is the volume of the crystal, and the energy splitting from valence to conduction band near the Dirac point is 
\be\label{eq:deltaE}
\Delta E_{\bk ,\bk'} = \sqrt{\tilde\bk^2+ \Delta^2} + \sqrt{\tilde\bk'^2+ \Delta^2}= \sqrt{\tilde\bk^2+ \Delta^2} + \sqrt{(\tilde\bk+\tilde\q)^2+ \Delta^2}\,,
\ee
with $\tilde\bk=(v_{Fx}k_x,v_{Fy}k_y,v_{Fz}k_z)$.

\begin{figure}[t]
\begin{center}
\includegraphics[width=1\textwidth]{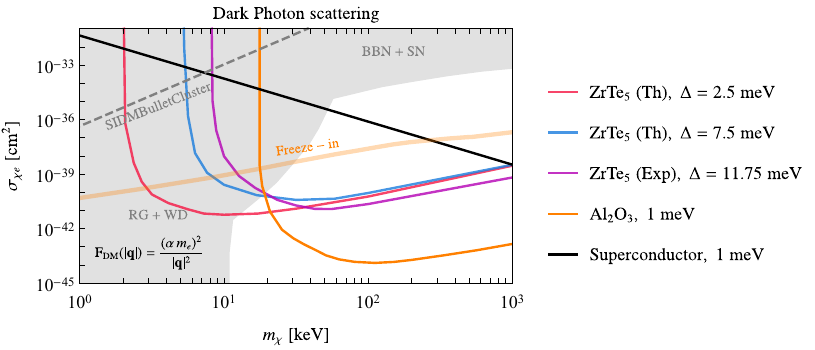}
\caption{\emph{Projected reach of dark matter scattering in Dirac material for a background-free $95\%$ C.L. sensitivity ($3$ events) assuming a  $1 \Kg$-year exposure of ${\rm ZrTe}_5$ with band gaps of $2.5\meV$ (red line), $7.5\meV$ (blue line) and $11.75\meV$ (purple line). For ${\rm ZrTe}_5$ curves, ``Th'' indicates the use of theoretical parameters, and ``Exp'' indicates the use of experimental parameters (See Table \ref{tab:zrte5values}). For comparison, we show the respective reaches of superconductors with a $1 \meV$ threshold (black line) \cite{Hochberg:2015fth} and sapphire ($Al_2O_3$) with a $1 \meV$ threshold (orange line) \cite{Knapen:2017ekk}. The thick orange line indicates the region of parameter space where the freeze-in production results in the correct dark matter relic abundance, as computed in Ref.~\cite{Dvorkin:2019zdi}. Shaded regions are bounds from red giants, white dwarfs, big bang nucleosynthesis and supernovae, and are derived from millicharged particle limits \cite{Essig:2015cda,Davidson:2000hf}. The dashed line is the self interacting dark matter bound derived from observations of the Bullet Cluster \cite{Feng:2009hw}.}}
\label{fig:scattreachplot}
\end{center}
\end{figure}

\begin{figure}[t]
\begin{center}
\includegraphics[width=\textwidth]{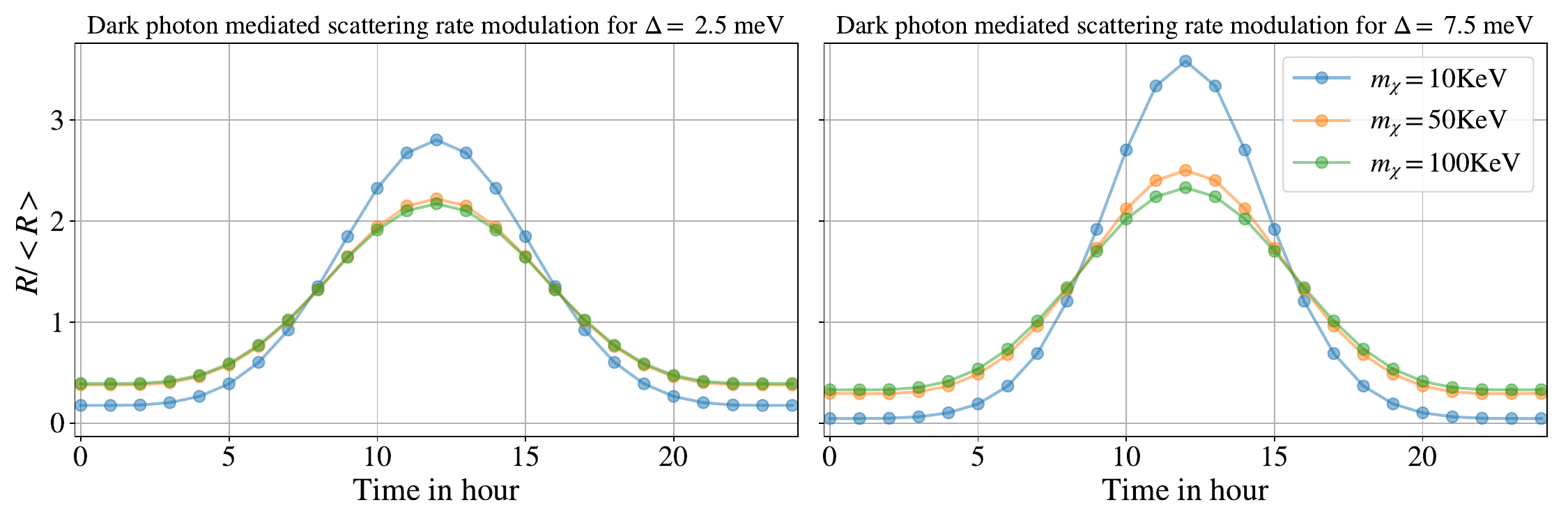}
\caption{\emph{Daily modulation of the dark photon mediated scattering rate for ${\rm ZrTe_5}$ (using theoretical parameters, see Table \ref{tab:zrte5values}) with a band gap of $2.5\meV$ (left panel) and $7.5\meV$ (right panel) for three different dark matter masses.}}
\label{fig:scattmod}
\end{center}
\end{figure}
We utilize a convention where $\overbar{\sigma}_e$ is the fiducial cross-section
\be
\overbar{\sigma}_e = \frac{16 \pi \mu_{\chi e}^2 \varepsilon^2 \alpha_{\textrm{EM}} \alpha_{\textrm{D}}}{q_0^4}, 
\ee
and $F_{\textrm{med}}(|\textbf{q}|)$ incorporates the momentum dependence in the scattering cross-section,
\be
F_{\textrm{med}}(|\textbf{q}|) = q_0^2/|\textbf{q}|^2,\ \  q_0 = \alpha m_e.
\ee
Because $\omega$ depends on ${\bf q } \cdot {\bf v}$, the energy conserving delta function can be used to evaluate the velocity integral as proposed in Ref.~\cite{Griffin:2018bjn}:
\beq\label{eq:gfunct}
\int d^3 v f_{\rm lab}({\bf v}) \delta(\Delta E - \omega ) =  \frac{\pi v_0^2}{N_0 |\q|} \left[\mbox{exp}(-v_-^2/v_0^2)-\mbox{exp}(-v_{\rm esc}^2/v_0^2)\right] \equiv g({\bf q},\Delta E_{\bk ,\bk'}),
\eeq
where $N_0$ is a normalization constant defined in \eq{eq:normcost}, and 
\beq
v_- = \mbox{min}\left\{\left|\frac{{\bf q} \cdot {\bf v}_e + \q^2/2 m_\chi + \Delta E}{|\textbf{q}|}\right|,v_{\rm esc}\right\}
\label{eq:vmin}
\eeq
is the minimum velocity (in the galactic reference frame) that a DM particle should have to induce a scattering with energy deposition $\Delta E$ and momentum transfer $\q$.
To obtain the total rate, $R_{-,\bf{k} \rightarrow +,\bf{k'}}$ is summed over initial and final state BZ momenta, over a region of size $\Lambda$ near the Dirac point
\be
R_{\rm crystal}= g_s V^2 \int_{\textrm{BZ}}  \frac{d^3 \bk d^3 \bk'}{(2 \pi)^6} R_{-,\bf{k} \rightarrow +,\bf{k'}} = g_s g_C V^2 \int_{\textrm{cone}}  \frac{d^3 \bk d^3 \bk'}{(2 \pi)^6} R_{\bf{k} ,\bf{k'}}\,.
\ee
The scattering rate per unit time per unit detector mass is then given by
\be\label{eq:scattrate}
R = \frac{\rho_\chi}{m_\chi} \frac{\pi \overbar{\sigma_e}}{ \mu_{\chi e}^2}  \int \frac{d^3 \textbf{q}}{(2\pi)^3}  \int \frac{d^3 \bk}{(2 \pi)^3} F_{\textrm{med}}(\q)^2\;  |\langle f | {\cal F}_T | i \rangle |^2\; g({\bf q},\Delta E_{\bk ,\bk'})
\ee
with
\be\label{eq:matelement}
 |\langle f | {\cal F}_T | i \rangle |^2 = \pi g \ n_e V_{\textrm{uc}} \underbrace{\left(\frac{\textbf{q}^2}{|\textbf{q} \cdot \boldsymbol{\epsilon} \cdot \textbf{q}|}\right)^2}_{\textrm{$\epsilon$-factor}} \times \underbrace{\left(1 -  \frac{\tilde{\bk} \cdot (\tilde{\bk}+\tilde{\textbf{q}})}{|\tilde{\bf{k}}| |\tilde{\bk}+\tilde{\textbf{q}}|} \right)}_{\textrm{WF-factor}},
\ee
where $\textit{V}_{\textrm{uc}}$ is the volume of the unit cell, $n_e$ is the electron density, and $g=g_s g_C$ is the product of Dirac cone and spin degeneracies. For future convenience, we labeled two factors in the Matrix Element $\sum_f |\langle f | {\cal F}_T | i \rangle |^2$ as $\epsilon$ and WF factors respectively.

In \Fig{fig:scattreachplot} we show the projected reach for a ${\rm ZrTe}_5$ target with gap values of $2.5\meV$, $7.5\meV$ and $11.75\meV$, and we compare it with the one of a superconducting target with $1 \meV$ threshold \cite{Hochberg:2016ajh}.  Because of the strong optical response of the superconductor \cite{Hochberg:2015pha,Hochberg:2015fth} compared to the Dirac Materials, Dirac Materials are far superior in reach. By contrast, polar materials \cite{Knapen:2017ekk} have a better reach. Note that our results again differ from those presented in \cite{Hochberg:2017wce} and \cite{Geilhufe:2019ndy}, where the dielectric constant was underestimated by an order of magnitude leading to an $\mathcal{O}(10^2)$ larger rate; a more detailed discussion of the differences between the present calculation and previous works is given in Appendix~C. 

In \Fig{fig:scattmod} we report the daily modulation of the scattering rate for ${\rm ZrTe}_5$ for different mass and threshold values. We observe that daily modulation increases for higher masses and thresholds. Also, the rate is maximized around $t=1/2$ day when the DM wind point along the smallest component of the dielectric tensor. The reason for these behaviors is discussed in detail next.

\subsection{Origin of the daily modulation}
The crystal anisotropies and the Earth's motion single out two preferred directions for the momentum transfer. The relative orientation of these directions changes during the day because of Earth's rotation giving rise to a daily modulation. Specifically, when these two directions coincide we have a maximum of the scattering rate and vice versa. 

Crystal anisotropies affects the rate in two ways. First, they appear through the anisotropy in the Fermi velocity; the kinematics prefers the momentum transfer to be anti-aligned with the direction of the smallest Fermi velocity, in order to increase the available phase space. We refer to this as a kinematic effect. And second they affect the magnitude of the scattering matrix element, $|\langle f | {\cal F}_T | i \rangle |$, through the so-called $\epsilon$ and ``WF'' factors labeled in \eq{eq:matelement}.  We now discuss the preferred direction singled out by each one of these two elements together with the Earth motion.

\subsubsection{Anisotropies in Kinematics}\label{anisotropykin}
 
\begin{figure}[t]
\begin{center}
\includegraphics[width=\textwidth]{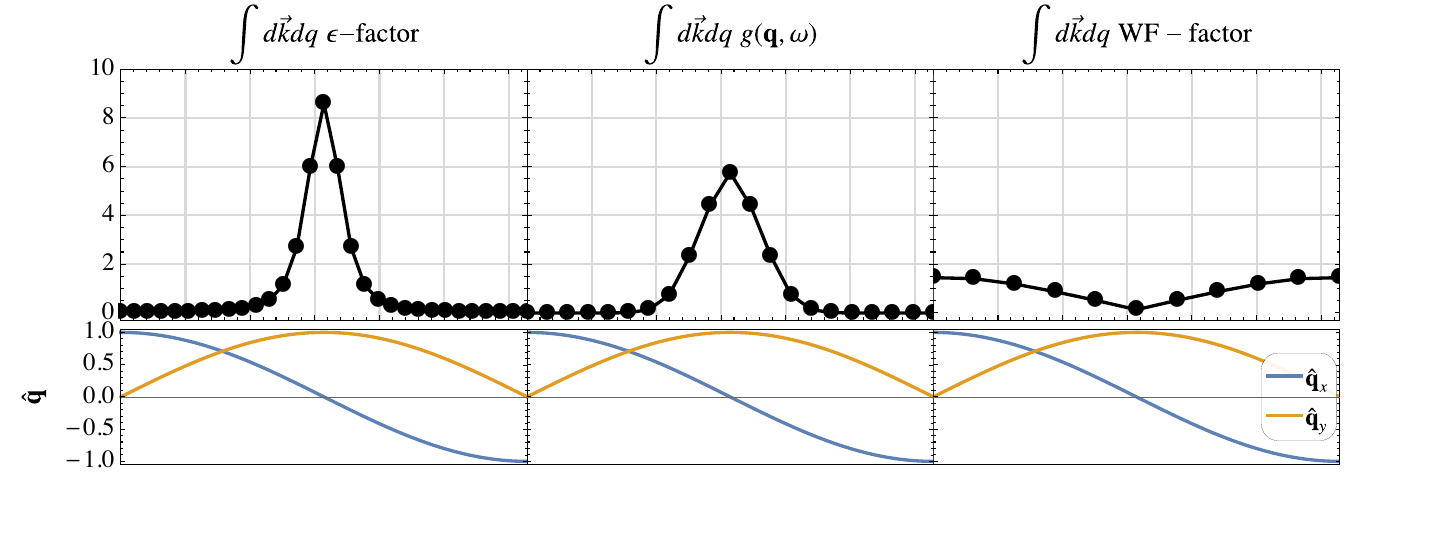}
\caption{\emph{Values of the $\epsilon$-factor (left panel), kinematic term (central panel), and WF-factor (right panel) for different orientation of $\q$ in the x-y plane (which is the one presenting the largest anisotropies). The reported values (which are normalized to its average value) are obtained by taking $m_\chi=50 \keV$, $\Delta=2.5\meV$, and  integrating over the magnitude of $\q$. To disentangle the effect of crystal anisotropies and Earth motion in the kinematic term, we have varied the orientation of $\q$ while keeping the angle between $\q$ and the Earth velocity fixed.}}
\label{fig:crystani}
\end{center}
\end{figure}
 
The kinematics of the scattering is sensitive to anisotropies in the Fermi velocity due to the rescaled momenta $\tilde\bk$ and $\tilde\q$ that appear in $\Delta E_{\bk,\bk'}$. From \eq{eq:deltaE} we can see that, for a given $\bk$, the energy transfer is minimized when $\q$ is aligned with the smallest component of the Fermi velocity. Therefore, when $\q$ points in this direction, a larger fraction of momenta in the BZ is kinematically allowed, so that the rate is maximized (as shown in the central panel of \Fig{fig:crystani}).

\subsubsection{Anisotropies in the Matrix Element}
 
\begin{figure}[t]
\begin{center}
\includegraphics[width=0.38\textwidth]{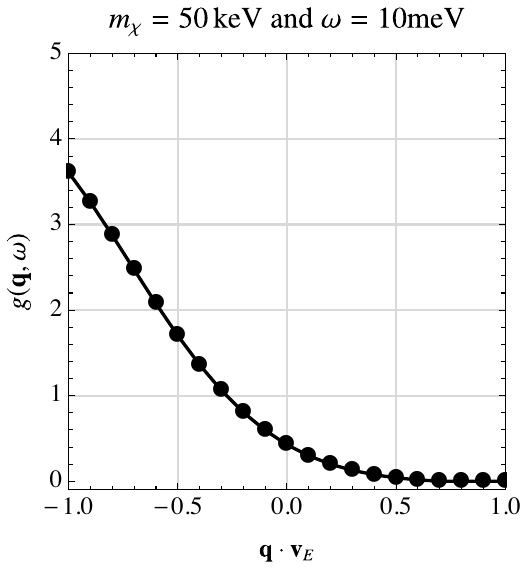}\qquad
\includegraphics[width=0.5\textwidth]{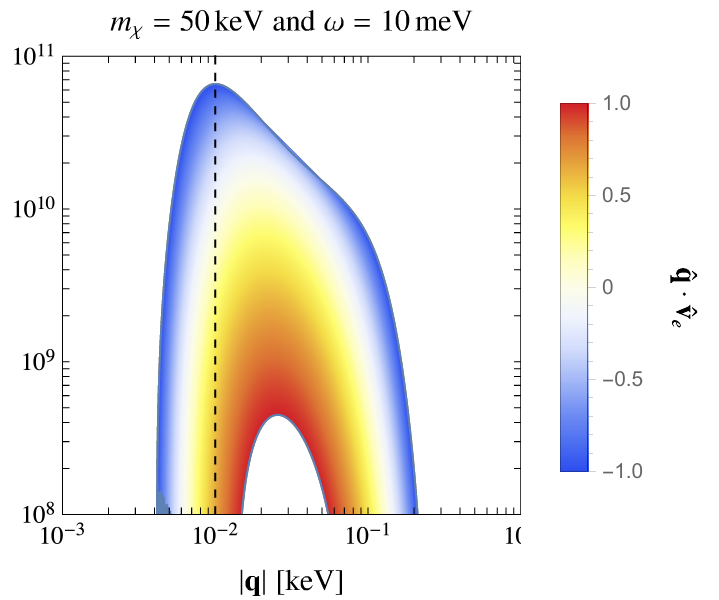}
\caption{\emph{{\bf Left}: Value of the kinematic factor $g(\q,\Delta E)$ (normalized to its average value) for different values of $\q\cdot\textbf{v}_e$ and a fixed value of $|\q|$. {\bf Right}: kinematic factor $g(\q,\Delta E)$ for different values of $|\q|$ and $\q\cdot\textbf{v}_e$. The dashed line indicates the value of $|\q|$ assumed in the left panel.}}
\label{fig:gplot}
\end{center}
\end{figure}
 
The scattering matrix element is sensitive to anisotropies in both the band structure and the dielectric tensor. To understand the modulation induced by the matrix element it is useful to deconstruct the impact on the matrix element in terms of the $\epsilon$ and WF factors defined in \eq{eq:matelement}. 

Crystal anisotropies enter in the $\epsilon$-factor thorough the tensorial structure of $\e$. Namely, the $\epsilon$-factor is maximized when $\q$ points in the direction of the smallest component of the dielectric tensor (as shown in the left panel of \Fig{fig:crystani}).

The WF-factor is sensitive to anisotropies in the band structure of the crystal through the rescaled momentum transfer $\tilde\q$. The WF-factor grows by increasing the angle between $\tilde\q$ and $\tilde\q+\tilde\bk$. This angle is non-zero when $\tilde\q$ gives a sizable contribution to the sum $\tilde\q+\tilde\bk$. The fraction of the kinematically allowed $\q$ that satisfies this requirement grows when the momentum transfer is aligned with the largest component of the Fermi velocity, given that this maximized the rescaled momenta $\tilde\q$. The WF-factor is maximized (minimized) when the momentum transfer points along the direction with the highest (smallest) Fermi velocity, as shown in the right panel of  \Fig{fig:crystani}.

\subsubsection{Earth motion}
The Earth's motion picks out a preferred direction for $\q$ through the minimum velocity, $v_-$, appearing in the velocity phase space integral \eq{eq:vmin}. A smaller minimum velocity corresponds to a larger available phase space, and hence a larger $g({\bf q},\Delta E_{{\bf k},{\bf k}'})$ and a larger rate.  As shown in \Fig{fig:gplot}, this function is maximized when $\q$ is antiparallel to $\bfv_e(t)$.  
\\
\\
We now have all the ingredients to understand the daily modulation shown in \Fig{fig:scattmod}. Because of the large anisotropies in the $\epsilon$-factor (see \Fig{fig:crystani}), the rate is dominated by momentum transfers along the $y$-direction, corresponding to the smallest component of the dielectric tensor. Therefore, the average value of the scattering rate is approximately given by fixing $\q\parallel \hat y$ in the integrand of \Eq{eq:scattrate}. A daily modulation around this average value is induced by the rotation of the Earth via the kinetic term. This term reaches is maximum when the Earth velocity is antiparallel to $\q$ (see \Fig{fig:gplot}). Because of this, the scattering rate is maximized at $t=1/2$ day, when the Earth velocity is aligned with the $y$-axes.

 \section{Conclusions}
Dirac materials, due to their small band gap, are promising targets to look for scatterings and absorption of light dark matter. In this work we discussed how crystal anisotropies can give rise to a daily modulation of the DM interaction rate with a Dirac material. 

To this end, we generalized the discussion of Ref.~\cite{Hochberg:2017wce} to accurately take into account anisotropies of the crystal structure, which induce a mixing of the longitudinal and transverse dark photon polarizations during their propagation in the media. We then applied these results to the specific case of the Dirac material ${\rm ZrTe}_5$ finding no modulation for the absorption signal and a daily modulation of $\mathcal{O}(1)$ for the scattering signal (see \Fig{fig:scattmod}).  In \Fig{fig:absreachplot} (\Fig{fig:scattreachplot} ) we also provide new projected reach for the absorption (scattering) signal, which is one (two) orders of magnitude weaker in comparison to Refs.~\cite{Hochberg:2017wce} and \cite{Geilhufe:2019ndy}. This discrepancy, as discussed in Appendix \ref{app:comparison}, is due to an incorrect prescription for incorporating the dielectric tensor, which underestimated the impact of in-medium screening.

Since, for the scattering signal, the shape of the modulation depends on the orientation of the crystal with respect to the dark matter wind, modulation effects can be used to validate any putative dark matter signal simply by changing the orientation of the crystal and observing a change in the shape of the modulation. Moreover, since the amplitude of the modulation depends on the DM mass, a definitive observation of the modulation could be used to infer the value of the DM mass. 

 \section*{Acknowledgments}
We thank Sinead Griffin, Adolfo Grushin, Tanner Trickle and Kevin Zhang for discussions, and Yoni Kahn for comments on the draft.  We especially thank Marco Bernardi and Hsiao-Yi Chen for discussions and collaboration on future related work. This work was supported by the Quantum Information Science Enabled Discovery (QuantISED) for High Energy Physics (KA2401032) at LBNL.
 \appendix
 
\section{In medium polarization tensor}\label{app:poltens} 
The in-medium vacuum polarization tensor can be related to the optical response of the medium by using the two following relations \cite{Schrieffer}:\footnote{To be precise the relation between $J$ and $\Pi$ is given by $J_\mu=-R_{\mu\nu}A^\nu$ and the imaginary part of $R$ and $\Pi$ have a different sign when $\omega$ is negative. This does not seems to be ever important for us.}
\begin{align} 
&J_\mu=-\Pi_{\mu\nu}A^\nu \label{eq:4J}\\
&J^i=\sigma^i_{\;j}E^j\label{eq:spatialJ}
\end{align}
where $\sigma_{ij}= i\omega(\delta_{ij}-\e_{ij})$ is the conductivity tensor. Specifically, using eq.\eqref{eq:spatialJ} and taking the spatial component of eq.\eqref{eq:4J} we get:
\be
\left\{\begin{array}{l}
J_i=-\Pi_{ij}A^j-\Pi_{i0}A^0\\ \\
J_i=\sigma_{ij}E^j=\sigma_{ij}\left(i\omega A^j-iq^jA_0\right)
\end{array}\right.
\ee
where in the second equation we have used the Maxwell equation $E^j=i\omega A^j-iq^j A_0$. From this we see that 
\begin{align}\label{eq:ptoe}
&\Pi_{ij}=-i\omega\sigma_{ij}\\
&\Pi_{i0}=i\sigma_{ij}q^j\,.
\end{align}
While from \Eq{eq:spatialJ}, together with current conservation $\partial _\mu J^\mu=0$ and Maxwell equation $E^j=i\omega A^j-iq^j A_0$, we get 
\be 
\Pi_{00}=\frac{1}{i\omega}\vec q\cdot\sigma\cdot\vec q\,.
\ee

\section{Dielectric tensor}\label{app:diel}
Following \cite{PhysRevB.73.045112}, the imaginary part of the macroscopic dielectric tensor is given by the Lindhard formula  
\be\label{eq:Imdiel}
\textrm{Im}[\e_{ii}(\omega)]=\frac{g e^2}{\q^2}\lim_{q\to0}\sum_{nn'}\int\frac{d^3\bk}{(2\pi)^3}2\pi\,\delta(E_{n'\bk}-E_{n\bk}-\omega)|f_{[n\bk\to n'\bk+q\,\hat{\mathbf{e}}_i]}|^2\,,
\ee
where $\hat{\mathbf{e}}_i$ are the unit vectors for the three cartesian components, and the sum runs over the energy levels. The energy conserving delta ensures that for small $\omega$ (\emph{i.e.}, $\omega\lesssim v_{F}\Lambda\sim\eV$) only transition between the valence ($n=-$) and conduction ($n'=+$) band near the Dirac point will contribute. The form factor $|f_{[n\bk\to n'\bk']}|^2$ for these low energy transitions can be computed analytically in Dirac materials \cite{Hochberg:2017wce}:
\be\label{eq:formfact}
|f_{[-\bk\to +\bk']}|^2=\frac{1}{2}\left(1-\frac{\tilde\bk\cdot\tilde\bk'+\Delta^2}{\sqrt{\tilde\bk^2+\Delta^2}\sqrt{\tilde\bk'^2+\Delta^2}}\right)\,.
\ee
where, as in the main text, $\tilde\bk=(v_{Fx}k_x,v_{Fy}k_y,v_{Fz}k_z)$. Plugging this expression in \Eq{eq:Imdiel} we are able to compute the imaginary part of the dielectric tensor for low energy deposition (\emph{i.e.}, $\omega\lesssim v_{F}\Lambda\sim\eV$).

The real and imaginary part of the dielectric are related by the Kramers-Kronig relation:
\be\label{eq:KKrel}
\textrm{Re}[\e_{ij}(\omega)]=1+\frac{2}{\pi}\mathcal{P}\int_0^\infty\frac{\textrm{Im}[\e_{ij}(\omega')]\omega'}{\omega'^2-\omega^2}d\omega'
\ee
where $\mathcal{P}$ denotes the principal part value. From this relation it is clear that the real part of the dielectric, even for small values of $\omega$, receives contributions also from transitions between states far away from the Dirac point. For these transitions the analytic expression of the form factor given in \Eq{eq:formfact} is no longer accurate and we have to resort to a density functional perturbation theory calculation. The result, that we will use as an input for our calculations, is shown in Table \ref{tab:zrte5values}.

\section{Comparison with previous results}\label{app:comparison}
In this appendix we highlight the reasons behind the discrepancy between our results and the ones presented in references \cite{Hochberg:2017wce} and \cite{Geilhufe:2019ndy}. 

In these two works the dielectric was computed analytically exploiting the analogy between QED and low energy electronic excitations of Dirac materials (see Eq. (B.3) in \cite{Hochberg:2017wce}, and Eq. (3.5) and (3.6) in \cite{Geilhufe:2019ndy}).\footnote{The polarization tensor computed in Eq. (3.5) and (3.6) of \cite{Geilhufe:2019ndy} can be related to the dielectric tensor by using \Eq{eq:ptoe}} The analytic expression for the dielectric obtained is
\be\label{eq:analitice}
\e=\mathbb{I}+\left(\begin{array}{ccc} 
\frac{v^2_{F,x}}{\kappa_{xx}} & 0 &0\\
0 & \frac{v^2_{F,y}}{\kappa_{yy}} & 0\\
0 & 0 & \frac{v^2_{F,z}}{\kappa_{zz}}
\end{array}\right)\Pi(\tilde q^2)\,,
\ee
where $\kappa_{ij}$ is the background dielectric and $\Pi(\tilde q^2)$ is computed analytically via the expression
 \be
 \label{eq:Pi}
\begin{aligned}
 \Pi(\tilde q^2)= &= \frac{e^2 g }{4 \pi^2\, v_{F,x}v_{F,y}v_{F,z}} \left[\int_0^1 dx \left \{x(1-x) \textrm{ln} \left|\frac{(2 \tilde{\Lambda})^2}{\Delta^2 -x(1-x)\tilde q^2)} \right|  \right \} \right.\\ &\left.+i \,\frac{\pi}{6 } \sqrt{1 - \frac{4 \Delta^2}{\tilde q^2}}\left(1 + \frac{2 \Delta^2}{\tilde q^2} \right) \Theta(\tilde q^2 - 4 \Delta^2)\right]\;,
\end{aligned}
 \ee 
with $\tilde \Lambda\equiv \Lambda\times {\rm max}\left(v_{F,x},v_{F,y},v_{F,z}\right)$ and $\tilde{\bf q}=(v_{Fx}q_x,v_{Fy}q_y,v_{Fz}q_z)$. 

This analytic expression fails to reproduce the needed form of the real and imaginary parts of the dielectric that enters into both scattering and absorption rates.  First, we note that \Eq{eq:Pi} only includes the contribution from states near the Dirac point.   This does not encompass most of the contribution to photon screening, which is dominated by electron states far from the Dirac cone. Thus for the real part of the dielectric used in calculating both absorption and scattering rates, we must employ instead the values computed from Density Functional Theory (DFT), shown in Table~I.  Second, the Lindhard formula \Eq{eq:Imdiel} encodes the relevant information of the imaginary part of the dielectric needed, in particular, for absorption rates; the imaginary part of \Eq{eq:analitice}, however, reproduces the Lindhard formula only in the limit that $\kappa_{ii} \rightarrow 1$.  We verified this prescription by comparison with DFT results.\footnote{We thank Hsiao-Yi Chen and Marco Bernardi for this comparison and for collaboration on future work.}  These two important differences summarize the changes in comparison to Refs.~\cite{Hochberg:2017wce} and \cite{Geilhufe:2019ndy}, where the absorption and scattering rates computed here differ by at least an order of magnitude in comparison to previous results.  

A procedure, akin to a renormalization and matching scheme, that manages to encapsulate the contribution of the high energy excitations in \Eq{eq:analitice} with $\kappa_{ii} \rightarrow 1$, and gives qualitatively the same results of the procedure used in the main text is the following.  The difference between the polarization tensor measured at two different energy scales (within the region of validity of the low energy effective theory) is finite and given by 
\be
\Pi(\tilde q^2)-\Pi(\tilde q_0^2)=\frac{e^2 g}{4\pi^2 v_{F,x}v_{F,y}v_{F,z}}\int_0^1dx\,x(1-x)\left[\ln\left(\frac{\Delta^2-\tilde q_0^2 x(1-x)}{\Delta^2-\tilde q^2 x(1-x)}\right)\right]\,,
\ee
by computing $\e(0,0)$ from DFT we can extract the value of  $\Pi(0)\simeq\e(0,0)/v_{\textrm{F}}^2$ and write 
\be\label{eq:norm2}
\begin{split}
\Pi(\tilde q^2)&=\Pi(0)+\frac{\tilde e^2 g}{4\pi^2 v_{F,x}v_{F,y}v_{F,z}}\left[\int_0^1dx\,x(1-x)\,\ln\left|\frac{\Delta^2}{\Delta^2-\tilde q^2 x(1-x)}\right|\right.\\
&\left.+i \,\frac{\pi}{6 } \sqrt{1 - \frac{4 \Delta^2}{\tilde q^2}}\left(1 + \frac{2 \Delta^2}{\tilde q^2} \right) \Theta(\tilde q^2 - 4 \Delta^2)\right]\;.
\end{split}
\ee
The real part of $\Pi(\tilde q^2)$ turns out to be dominated by $\Pi(0)$, suggesting that the main contribution to the dielectric comes from electron states far from the Dirac cone. Moreover, this justifies the assumption made in the main text, \emph{i.e.} $\e(\omega,\q)\simeq\e(0,0)$. By contrast, since $\Pi(0)$ is real, the imaginary part of \Eq{eq:norm2} is the same of \Eq{eq:Pi}. 

Finally, the anisotropic dielectric tensor derived in Appendix B of Ref.~\cite{Hochberg:2017wce} carries a $\tilde\q^2$-dependence which would lead to a daily modulation for the absorption rate. As already pointed out by Ref.~\cite{Geilhufe:2019ndy}, the $\tilde\q^2$ dependence in Eq.~(B.3) of Ref.~\cite{Hochberg:2017wce} derives from an erroneous generalization of the isotropic dielectric; the correct expression replaces $\tilde\q^2 \rightarrow \q^2$ in Eq.~(B.3) of Ref.~\cite{Hochberg:2017wce} and implies no daily modulation rate for absorption. 

\bibliography{bibliography.bib}

\end{document}